\documentclass[twocolumn,superscriptaddress,nopacs,aps,pra]{revtex4-2}

\usepackage{graphicx}  
\usepackage{dcolumn} 
\usepackage{bm}  
\usepackage{amsmath}
\usepackage{amsthm}
\usepackage{amssymb}
\usepackage{comment}
\usepackage{xcolor}

\usepackage[T1]{fontenc}
\usepackage[utf8]{inputenc}
\usepackage[normalem]{ulem}
\usepackage{empheq}

\newcommand*\Eperp{\mathbf{E}_\perp}
\newcommand*\Laplaceperp{\nabla^2_\perp}
\newcommand*\nablaperp{\nabla_\perp}
\newcommand*\itot{I_0}
\newcommand*\Eperphat{\mathbf{\hat E}_\perp}
\newcommand*\nablaperphat{\hat\nabla_\perp}

\newcommand{\bs}[1]{{\boldsymbol{#1}}}

\newcommand{\bq}{\bs{q}}
\newcommand{\br}{\bs{r}}

\begin{document}

\title{Spin-orbit-induced gauge field and spin Hall effects in quasi-paraxial fluids of light}

\author{Caroline Stier}
\email{caroline.stier@physik.uni-freiburg.de}
\affiliation{%
Physikalisches Institut, Albert-Ludwigs-Universit\"at Freiburg,
Hermann-Herder-Straße 3, D-79104 Freiburg, Germany
}%

\author{Nicolas Cherroret}
\email{nicolas.cherroret@lkb.upmc.fr}
\affiliation{%
Laboratoire Kastler Brossel, Sorbonne Université, CNRS, ENS-PSL Research University, Collège de France,
4 Place Jussieu, 75005 Paris, France
}


\begin{abstract}
We show that quasi-paraxial fluids of light propagating in nonlinear Kerr media naturally experience an effective gauge field arising from the intrinsic spin-orbit interaction of light encoded in Maxwell equations. By deriving the leading nonparaxial corrections to the nonlinear Schr\"odinger equation governing the fluid evolution, we find that, for circularly polarized beams, these corrections generate an effective Lorentz force that modifies the fluid dynamics. Remarkably, the corresponding gauge field is not externally imposed but self-consistently generated by the fluid itself, with its sign controlled by the light helicity and its magnitude governed by the spatial variations of the nonlinear refractive index. We illustrate the physical consequences of this gauge field in two configurations: the chiral deformation of normal and superfluid flows past an obstacle, accompanied by a transverse spin Hall force, and the cumulative precession of an optical wave packet trapped in a focusing fluid.

\end{abstract}

\maketitle

\section{Introduction}

At the intersection of  nonlinear optics and  quantum many-body physics, the concept of fluid of light has proven particularly fruitful for exploring analogies and differences  between degenerate quantum gases and systems of effectively interacting photons \cite{Carusotto2013}. Among fluid-of-light platforms, paraxial beams propagating in nonlinear media are especially appealing, as their evolution closely resembles the conservative mean-field dynamics of degenerate Bose gases typically realized in ultracold-atom experiments \cite{Glorieux2025}. This analogy relies on the mathematical structure of the nonlinear paraxial wave equation, which describes the evolution of a two-dimensional transverse optical field along a longitudinal axis that plays the role of an effective time, with diffraction giving the photons an effective mass. 
In the presence of a Kerr nonlinearity, the photons experience effective contact interactions, and
the paraxial wave equation becomes formally identical to the Gross-Pitaevskii equation for dilute Bose gases. As such, this system has been shown both theoretically and experimentally  to support many phenomena familiar from quantum gases, including characteristic Bogoliubov dispersion \cite{Vocke2015, Fontaine2018}, the generation of superfluid flows past obstacles \cite{Frisch1992, Pomeau1993, Josserand1999, Pavloff2002, Astrakharchik2004, Pinsker2014, Vocke2016, Huynh2024, Michel2018, Rasooli2026, Larre2026},  topological defects such as vortices or solitons \cite{Coullet1989, Barad1997, Rasooli2026, Baker-Rasooli2025}, or the practical realization of quantum-quench problems with light \cite{Larré2018, Sun2012, Santic2018, Bardon-Brun2020, Abuzarli2022, Cherroret2024} to give a few examples.

While the polarization degree of freedom of paraxial fluids of light has long been overlooked, it has  recently been shown  that it could also be exploited to realize optical analogues of binary Bose gases \cite{Piekarski2025, Larre2026}, with the two circular-polarization components playing the role of spin-up and spin-down components of coupled Bose-Einstein condensates \cite{Pitaevskii2016, Recati2022}. This naturally raises the question of whether the ``spin'' degree of freedom of paraxial fluids can  also be used as a source of  \emph{spin-orbit interactions}, analogous to those arising from intrinsic or structural symmetries in condensed-matter systems \cite{Manchon2015, Sarma2024} and optical microcavities \cite{Kavokin2005, Leyder2007}, or synthetically engineered in cold-atom gases \cite{Dalibard2011, Lin2011, Galitski2013}. This is the question addressed in the present work.

Recent works \cite{Martone2021, Martone2023} have suggested that spin-orbit interactions naturally emerge for optical beams propagating in Kerr media, as a consequence of the intrinsic coupling between the polarization and the spatial variations of the electric field. 
In fact, this mechanism is known to naturally arise in any optical system involving a spatially-varying permittivity $\varepsilon(\br)$, as follows from Maxwell's equation $\nabla(\varepsilon(\br)\textbf{E}(\br))=0$ \cite{Liberman1992, Bliokh2015, Cardano2015, Aiello2015}. Spin-orbit interactions of light have thus been studied in a variety of settings, including reflection at interfaces \cite{Hosten2008, Qin2009, Gorodetski2012}, propagation in optical fibers \cite{Tomita1986, Bliokh2008} or disordered media \cite{Bardon-brun2019, Carlini2022, Carlini2023}.

In the context of fluids of light  in Kerr media, a similar  coupling arises from the spatial variations of the \emph{nonlinear} permittivity, which are self-induced by the variations of the beam intensity profile. In Refs.~\cite{Martone2021, Martone2023}, signatures of such spin-orbit coupling were identified  in the excitation spectrum of the fluid through a direct perturbative (Bogoliubov) treatment  of Maxwell equations. This approach, however, was restricted to weak perturbations around a homogeneous background and is not readily generalizable to more complex configurations. Moreover, spin-orbit interactions were only inferred indirectly, through their effect on a specific observable. This motivates the development of a more general framework for this physics, offering intuitive physical insight and applicability to a broad range of configurations.

In this paper, we develop such a framework by explicitly showing that, for fluids of light in Kerr media, the spin-orbit interactions intrinsically encoded in Maxwell's equations give rise to an effective gauge field in the paraxial wave equation. This gauge field naturally emerges from the leading nonparaxial corrections, which we derive through a systematic expansion around the paraxial limit (Sec. \ref{Sec:derivation}). The resulting spin-orbit-induced gauge field generates an effective Lorentz force, analogous to that experienced by a charged particle in a magnetic field. Its magnitude, however, is not externally imposed but self-consistently generated by the fluid itself: it is proportional to the light helicity, the strength of the nonlinearity, and the spatial variations of the fluid intensity profile. To illustrate the physical consequences of this gauge field, we highlight two direct manifestations. First, for a defocusing fluid of light, we show that it induces a chiral deformation of the intensity profile of both normal and superfluid flows past an index obstacle, and gives rise to a transverse force exerted on the obstacle (Sec. \ref{Sec:chiralflows}). Second, for a focusing fluid, we demonstrate that it leads to the cumulative precession of an optical wave packet trapped by a nonlinear background beam (Sec. \ref{sec:precession}). 
Beyond these examples, our approach is general and can be applied to a broad range of configurations and regimes, without assumptions on the fluid's profile or the magnitude of the nonlinearity.

\section{Effective gauge field in quasi-paraxial fluids of light}
\label{Sec:derivation}

\subsection{Helmholtz equation in a Kerr medium}

We study the evolution of a monochromatic laser beam in a non-magnetic, nonlinear dielectric and source-free medium. Writing the laser field as $\mathbf{E}(\br,t)={\rm Re}\left[\mathbf{E}(\br)e^{-i\omega t}\right]$, the amplitude $\mathbf{E}(\br)$ obeys the wave equation
\begin{align}
    \nabla^2\mathbf{E}-\nabla(\nabla\cdot\mathbf{E})=-\omega^2\mu_0\mathbf{D}.
    \label{waveeqD}
\end{align}
In a nonlinear medium, the electric displacement field $\mathbf{D}=\epsilon_0 n_0^2\mathbf{E}+\mathbf{P}^{\rm NL}$ can be  split  into a linear part, encoded in the background refractive index $n_0$, and a nonlinear (Kerr) part, $\mathbf{P}^{\rm NL}$. 
The most general third-order polarization is $\smash{P_i^{\rm NL}=\epsilon_0\chi^{(3)}_{ijkl}E_jE_kE_l^*}$, where indices $i=x,y,z$. For an isotropic Kerr medium (no preferred direction, as in an atomic vapor), $\smash{\chi^{(3)}_{ijkl}}$ must be built solely from delta tensors \cite{Agrawal2019, Boyd2020},
\begin{align}  \chi^{(3)}_{ijkl}=\alpha\,\delta_{ij}\delta_{kl}+\beta\,\delta_{ik}\delta_{jl}+\gamma\,\delta_{il}\delta_{jk}.
\end{align}
Since the two field factors $E_jE_k$ enter symmetrically, the first two terms contract to the same structure, so the nonlinear polarization takes the form
\begin{align}
    \mathbf{P}^{\rm NL}=\epsilon_0\left[(\alpha+\beta)\,|\mathbf{E}|^2\mathbf{E}+\gamma\,(\mathbf{E}\cdot\mathbf{E})\mathbf{E}^*\right].
    \label{PNLisotropic}
\end{align}
The first term is the familiar self-phase-modulation nonlinearity, proportional to the total intensity $|\mathbf{E}|^2$; the second term, proportional to $\mathbf{E}^*$ rather than $\mathbf{E}$, describes a coherent, degenerate four-wave-mixing process coupling $\mathbf{E}$ and $\mathbf{E}^*$. 
Substituting the isotropic form \eqref{PNLisotropic} into Eq.~\eqref{waveeqD}, and introducing  $k=\omega n_0/c$, $g=-k(\alpha+\beta)/2n_0^2$ and $g'=-k\gamma/2n_0^2$, we obtain
\begin{align}
    \nabla^2\mathbf{E}\!-\!\nabla(\nabla\cdot\mathbf{E})\!+\!k^2\mathbf{E}\!-\!2k(g|\mathbf{E}|^2\mathbf{E}\!+\!g'\mathbf{E}^2\mathbf{E}^*)\!=\!0.
    \label{helmholtz}
\end{align}
The physical significance of the two nonlinear constants $g$ and $g'$ in the language of fluids of light will be discussed below.

\subsection{Transverse and longitudinal fields}

At this stage, our goal is  to simplify Eq.~\eqref{helmholtz} so as to retain only the leading-order corrections beyond the paraxial limit. To this end, 
we  assume that the field propagates predominantly along $z$, and write
\begin{align}
    \mathbf{E}(\br)=\exp{\left(ikz\right)}\left[\mathbf{E}_\perp(\br_\perp,z)+\mathbf{e}_zE_z(\br_\perp,z)\right],
    \label{FormofE}
\end{align}
where $|E_z|\ll|\mathbf{E}_\perp|$ and $\br_\perp=(x,y)$. 
Inserting Eq.~\eqref{FormofE} into the Helmholtz equation \eqref{helmholtz} and projecting separately onto the transverse and longitudinal directions yields two coupled equations for $\mathbf{E}_\perp$ and $E_z$ \cite{Liberman1992}. The projection onto the transverse plane, first,  yields
\begin{align}
    i\partial_z\Eperp{}
    &=-\frac{(\Laplaceperp{}+\partial_z^2)\Eperp{}}{2k} 
    +\frac{\nablaperp\left(\nablaperp\!\cdot\!\mathbf{E}_\perp+(\partial_z+ik)E_z\right)}{2k}
    \nonumber\\
    &   +g(|\mathbf{E}_\perp|^2+|E_z|^2)\mathbf{E}_\perp+g'(\mathbf{E}_\perp^2+E_z^2)\mathbf{E}^*_\perp,
    \label{Etransvers}
\end{align}
where $\Laplaceperp{}=\partial_x^2+\partial_y^2$. 
This equation still depends on $E_z$, so it is not yet closed. Projecting instead onto $\mathbf{e}_z$
gives an equation for $E_z$, sourced entirely by the transverse field:
\begin{align}
    &\nablaperp^2E_z-ik\nablaperp\cdot\Eperp-\partial_z(\nablaperp\cdot\Eperp)+k^2E_z\nonumber\\
    &-2kg(|\Eperp{}|^2+|E_z|^2)E_z-2kg'(\mathbf{E}^2_\perp+E_z^2) E_z^*=0.
    \label{Ez}
\end{align}
Equation \eqref{Ez} shows that $E_z$ is not an independent dynamical degree of freedom: it is slaved to $\mathbf{E}_\perp$ and can, in principle, be eliminated by solving \eqref{Ez} for $E_z$ and substituting the result into \eqref{Etransvers}.

\subsection{Paraxial limit and leading nonparaxial corrections}
\label{Sec:paraxial}

From Eqs.~\eqref{Etransvers} and \eqref{Ez}, the paraxial limit is obtained by completely neglecting $E_z$, as well as the off-diagonal term $\propto\nabla_\perp(\nabla_\perp\cdot\mathbf{E}_\perp)$ and the second derivative $\partial_z^2\mathbf{E}_\perp$ in Eq.~\eqref{Etransvers}. Within this approximation, it is convenient to introduce the dimensionless variables
\begin{align}
\label{eq:dimensionless}
    \hat x=\frac{x}{\xi}, \qquad
    \hat y=\frac{y}{\xi}, \qquad
    \hat z=zg\itot, \qquad
    \mathbf{\hat E}=\frac{\mathbf{E}}{\sqrt{\itot}},
\end{align}
where $\itot$ denotes the characteristic intensity scale of the laser, and $\xi\equiv 1/\sqrt{2kg\itot}$ is the healing length, which measures the typical spatial variations of the paraxial beam due to the nonlinearity. In terms of these variables, the paraxial wave equation follows directly from Eq.~\eqref{Etransvers}:
\begin{align}
    i\partial_{\hat{z}}\Eperphat\simeq -\nablaperphat^2\Eperphat+|\mathbf{\hat E}_\perp|^2\mathbf{\hat E}_\perp+\eta\mathbf{\hat E}_\perp^2\mathbf{\hat E}^*_\perp
    \label{paraxialvectoreq},
\end{align}
where we have introduced the dimensionless parameter $\eta\equiv g'/g$. This equation closely resembles the nonlinear Schr\"odinger equation governing the dynamics of ultracold Bose gases, thereby justifying the concept of paraxial fluid of light \cite{Glorieux2025}. It is nevertheless important to note that Eq.~\eqref{paraxialvectoreq} still involves two field components, $\smash{{\hat E}_\perp^x}$ and $\smash{{\hat E}_\perp^y}$, which are generally coupled through the last term on the right-hand side. Only for linearly polarized light, for which $\mathbf{\hat E}_\perp=\mathbf{\hat E}^*_\perp$, and circularly polarized light, for which $\smash{\mathbf{\hat E}_\perp^2=0}$, does Eq.~\eqref{paraxialvectoreq} reduce to a single scalar wave equation. In all other cases, Eq.~\eqref{paraxialvectoreq} describes two coupled scalar equations.
When expressed in the circular-polarization basis, these equations become analogous to the coupled nonlinear Schr\"odinger equations describing mixtures of Bose gases, as discussed in \cite{Piekarski2025, Larre2026}. Within this analogy, the coupling constant $g$ describes the self-interaction of a given circular-polarization component, while $g_{12}=2g'+g$ characterizes the cross-coupling between the two components.

The usefulness of introducing the dimensionless variables defined above becomes particularly apparent from Eq.~\eqref{paraxialvectoreq}: apart from the parameter $\eta$, the dimensionless paraxial wave equation contains only unit coefficients. This provides a convenient framework for identifying the leading-order corrections to the paraxial approximation. Specifically, these corrections correspond to all contributions from Eqs.~\eqref{Etransvers} and \eqref{Ez} that are \emph{not} associated with either unit or $\propto\eta$ prefactors. 
To identify these corrections systematically, we  first rewrite Eqs.~\eqref{Etransvers} and \eqref{Ez} in terms of the dimensionless variables in Eq.~\eqref{eq:dimensionless}. This yields two exact, coupled dimensionless equations for the rescaled fields $\smash{\hat{\mathbf{E}}_\perp}$ and $\hat{E}_z$, in which the nonparaxial terms appear as powers of the dimensionless parameter
\begin{equation}
\epsilon\equiv (k\xi)^{-1}\ll1.
\end{equation}
This parameter is the ratio of the optical wavelength to the healing length $\xi$, and naturally controls all deviations from the paraxial limit. 
To derive a single effective Schr\"odinger equation for $\Eperphat$ that incorporates the leading nonparaxial corrections, we proceed as follows. 
We first use the dimensionless equation for the longitudinal field to express $\hat{E}_z$ as a function of $\smash{\Eperphat}$. We then  substitute this expression into the dimensionless equation for the transverse field and retain only the leading-order terms in $\epsilon$. The details of this calculation are provided in Appendix \ref{appendixA}. The resulting equation reads
\begin{align}
    i\partial_{\hat z}\Eperphat&=-\nablaperphat^2\Eperphat+|\mathbf{\hat E}_\perp|^2\mathbf{\hat E}_\perp+\eta\mathbf{\hat E}_\perp^2\mathbf{\hat E}^*_\perp-\frac{\epsilon^2}{4}\partial_{\hat z}^2\Eperphat{}\nonumber\\
    &+\epsilon^2\nablaperphat(\Eperphat\cdot\nablaperphat|\Eperphat|^2)+\eta\epsilon^2\nablaperphat(\Eperphat^*\cdot\nablaperphat\Eperphat^2)\nonumber\\
    &\!\!\!\!\phantom{\frac{1}{1}}+2\eta\epsilon^2\nablaperphat(\Eperphat^2\nablaperphat\cdot\Eperphat^*)+\mathcal{O}(\epsilon^4).
\label{Etransversfinal}
\end{align}
This equation describes the evolution of the transverse field along $z$, including the leading nonparaxial corrections, which appear at order $\epsilon^2$. 
Note that the absence of  corrections  of order $\epsilon$ follows from the rotational invariance of the transverse plane. This can already be seen from the three-dimensional dispersion relation $k_z=\sqrt{k^2-q_\perp^2}$, whose Taylor expansion in $q_\perp$ necessarily involves even powers of $q_\perp\sim 1/\xi$. The first line of Eq.~\eqref{Etransversfinal} reproduces the paraxial equation \eqref{paraxialvectoreq}, supplemented by the diffraction-order correction $\smash{\propto\partial_{\hat z}^2\Eperphat}$. The last two lines are particularly interesting: they involve nonparaxial terms that naturally encode a mechanism of  \emph{spin-orbit interaction} of light, by coupling the spatial structure of the paraxial-field envelope to its polarization. 

As a check of the validity of our approach, we have verified that the evolution equation \eqref{Etransversfinal} exactly reproduces the anisotropic sound velocities obtained in Ref. \cite{Martone2021} from the full Maxwell equations, in the particular example of a small field perturbation propagating on top of a homogeneous background.

\subsection{Spin-orbit induced gauge field for circularly polarized fluids}

To further elucidate the physical content of Eq.~\eqref{Etransversfinal}, we now specialize it to the particular case of a purely circularly polarized field, $\smash{\textbf{E}_\perp(\br_\perp,z)=\psi(\br_\perp,z) \,\textbf{e}_\sigma}$, where $\textbf{e}_\sigma=(\textbf{e}_x+i\sigma\textbf{e}_y)/\sqrt{2}$ and $\sigma=\pm1$ denotes the helicity of the beam. Projecting Eq.~\eqref{Etransversfinal} onto $\textbf{e}_\sigma^*$, using the fact that $\mathbf{\hat E}_\perp^2=0$ for circularly polarized light, and restoring dimensionful units, we obtain
\begin{align}
    i{\partial_z\psi}&=-\frac{\nabla_\perp^2\psi}{2k}
    +g|\psi|^2\psi-\frac{\partial_z^2\psi}{2k}+\frac{g}{k^2}D_\sigma(\psi D_{-\sigma} I)
    \label{eq:GPEcirc}
\end{align}
where $I \equiv |\psi|^2$ and $D_\sigma\equiv (\partial_x-i\sigma\partial_y)/\sqrt{2}$. The last term on the right-hand side can be simplified using
\begin{align}
        D_\sigma(\psi D_{-\sigma} I)=&
    \frac{1}{2}\nabla_\perp\cdot(\psi\nabla_\perp I)\nonumber\\
    &+\frac{i\sigma}{2}(\nabla_\perp I\times \textbf{e}_z)\cdot\nabla_\perp\psi.
\end{align}
Absorbing the second term on the right-hand side of this relation into the kinetic contribution of Eq.~\eqref{eq:GPEcirc}, we finally obtain:
\begin{align}
    i{\partial_z\psi}=&
    -\frac{(\nabla_\perp-\!ik\boldsymbol{A})^2\psi}{2k}
    +g|\psi|^2\psi\nonumber\\
    &\ \ \ \ \ -\frac{\partial_z^2\psi}{2k}+\frac{g}{2k^2}\nabla_\perp\cdot(\psi \nabla_\perp I)
    \label{eq:GPEgauge},
\end{align}
where $\boldsymbol{A}\equiv g\sigma/(2k^2)(\nabla_\perp I\times \textbf{e}_z)$, and where we have neglected the term proportional to $\boldsymbol{A}^2$, which is of order $\mathcal{O}(\epsilon^4)$.
Equation \eqref{eq:GPEgauge} constitutes the main result of the paper. Physically, it shows that a circularly polarized fluid of light experiences the leading nonparaxial correction as an effective Lorentz force, exactly as a charged particle would in a synthetic magnetic field $\boldsymbol{B}=(\nabla_\perp\times \boldsymbol{A})\cdot\textbf{e}_z$. 
The strength of this field is not fixed externally, but is generated self-consistently by the fluid itself: it is proportional to the local gradient of the intensity profile, $g\nabla_\perp I$, and its sign is set by the beam helicity $\sigma$. 
In regions where the fluid is spatially uniform, $\boldsymbol{A}$ vanishes and the ordinary paraxial dynamics is recovered. Wherever the intensity varies, for instance around a defect or at the edge of an inhomogeneous beam, the two circular polarizations experience opposite effective Lorentz forces. This is the optical analog of a spin-orbit coupling. The polarization (spin) becomes dynamically coupled to the orbital, spatial degrees of freedom of the fluid. In the next two sections, we will illustrate two different manifestations of this coupling: a chiral deformation of the flow pattern around a fixed obstacle (Sec. \ref{Sec:chiralflows}), and a slow, cumulative precession of a wave packet trapped in a strong Gaussian beam (Sec. \ref{sec:precession}).

The last two terms in Eq.~\eqref{eq:GPEgauge} are also nonparaxial contributions, but they provide small, helicity-independent corrections to the fluid’s dispersion, which will be neglected in the rest of this work.

\section{Application 1: chiral optical flows past an obstacle and spin Hall force}
\label{Sec:chiralflows}

\subsection{Optical flow past a weak-amplitude obstacle}

As a first practical illustration of the impact of the effective gauge field on a fluid of light, we consider the Landau problem of an optical fluid flowing past an obstacle. This question has been previously  addressed in a number of theoretical works \cite{Pavloff2002, Astrakharchik2004, Frisch1992, Pomeau1993, Josserand1999, Pinsker2014, Huynh2024}, 
and recently investigated experimentally using paraxial fluids of light in nonlinear crystals \cite{Michel2018} and atomic vapors \cite{Rasooli2026}. To this end, we consider a defocusing fluid of light ($g>0$) and add a stationary, localized potential $V(\br_\perp)$ into the nonlinear Schr\"odinger equation \eqref{eq:GPEgauge}:
\begin{align}
    i{\partial_z\psi}=&
    -\frac{(\nabla_\perp-\!ik\boldsymbol{A})^2\psi}{2k}
    +g|\psi|^2\psi+V(\br_\perp)\psi
    \label{eq:GPEgauge_V}.
\end{align}
In experiments involving paraxial fluids, the $z$-independent potential $V$ can be generated using an auxiliary focused beam, whose cross-phase nonlinearity is exactly compensated by diffraction \cite{Michel2018, Rasooli2026}.

To examine the impact of the spin-orbit gauge field on the fluid flow around the defect, we restrict our analysis to the weak-obstacle regime, $V\ll g|\psi|^2$, where standard perturbation theory can be applied to the field. In the absence of the obstacle, we assume that the beam is prepared in a tilted plane-wave state,
$\sqrt{I_0}e^{i q_0 x-i\mu z}$, with $\mu=q_0^2/(2k)+gI_0$, corresponding to a fluid carrying a finite momentum $q_0$ along the $x$ direction \cite{Rasooli2026}. In the presence of the obstacle, we seek a perturbative solution of Eq.~\eqref{eq:GPEgauge_V} of the form
\begin{equation}
\label{eq:linearization2}
\psi(\br_\perp,z)=e^{i q_0 x -i\mu z}[\sqrt{I_0}+\delta\phi(\br_\perp)],
\end{equation}
where $\delta\phi=\mathcal{O}(V)$. We focus here on the long-$z$ regime, in which the perturbation $\delta\phi(\br_\perp)$ has reached a stationary state and is therefore independent of $z$ \cite{Huynh2024}. 
Substituting this ansatz into Eq.~\eqref{eq:GPEgauge_V} and linearizing to first order in $\delta\phi$, we obtain the Bogoliubov-de Gennes equation
\begin{align}
\label{eq:BG_source}
    -\frac{\nabla_\perp^2\phi}{2k}
&-\frac{i q_0\partial_x\phi}{k}
+gI_0(\phi+\phi^*)
\nonumber\\
&-\frac{\sigma g I_0 q_0}{2k^2}
\partial_y (\phi+\phi^*)=-V\sqrt{I_0}.
\end{align}
Introducing the Fourier-space representation
$\delta\phi(\bq)=\int d^2\br_\perp,e^{-i\bq\cdot\br_\perp}\delta\phi(\br_\perp)$, Eq.~\eqref{eq:BG_source} can be recast as an algebraic Bogoliubov system for the vector $[\delta\phi(\bq),\delta\phi^*(\bq)]^T$, with source term $[-V(\bq)\sqrt{I_0},-V(\bq)\sqrt{I_0}]^T$. Solving this system yields
\begin{equation}
\label{eq:soldeltaphi}
    \delta \phi(\bq)=
    -\sqrt{I_0}
    \frac{V(\bq)}{\Delta(\bq)}
    \left(\frac{q^2}{2k}-\frac{q_0 q_x}{k}\right),
\end{equation}
with 
\begin{equation}
\label{eq:deltaform}
    \Delta(\bq)=
    \frac{q^2}{2k}\left(\frac{q^2}{2k}+2gI_0\right)
    -\frac{q_0^2 q_x^2}{k^2}
    -\frac{i\sigma g I_0 q_0}{2k^3}q^2 q_y
\end{equation}
and $\delta\phi^*(\bq)=-\delta\phi(\bq)$. The three terms on the right-hand side of this expression correspond, respectively, to the squared bare Bogoliubov dispersion, the Doppler shift associated with the finite fluid velocity, and the spin-orbit correction of interest here.
From Eq.~\eqref{eq:soldeltaphi}, the spatial intensity pattern of the fluid around the obstacle follows as $I(\br_\perp)=|\psi(\br_\perp)|^2=I_0+\delta I(\br_\perp)$, where 
\begin{align}
\label{eq:deltaI_interm}
    \delta I(\br_\perp)&=2\sqrt{I_0}\text{Re}[\phi(\br_\perp)]
    \nonumber\\
    &=-I_0\int \frac{d^2\bq}{(2\pi)^2}\frac{\bq^2}{k}\frac{V(\bq)}{\Delta(\bq)} e^{i\bq\cdot\br_\perp}.
\end{align}
Using Eq.~\eqref{eq:deltaform}, the intensity distribution can be rewritten as
\begin{align}
\label{eq:Intensitypattern}
\frac{ \delta I(\br_\perp)}{I_0}
\!=\!
\int\! \frac{d^2\bq}{(2\pi)^2}
\frac{-4kV(\bq)e^{i\bq\cdot\br_\perp}}{q^2\!-\!\frac{2}{\xi^2}(M^2 q^2/q_x^2\!-\!1)
\!-\!i\frac{\sigma M q_y}{\sqrt{2}k\xi^3}}
\end{align}
where we have introduced the Mach number $M\equiv q_0/\sqrt{gI_0k}=\sqrt{2}q_0\xi$.

It is instructive to examine the structure of Eq.~\eqref{eq:Intensitypattern}. In the absence of spin-orbit coupling, i.e., when the last term in the denominator vanishes, the denominator has real poles for $M>1$, i.e., $q_0/k>c_s\equiv \sqrt{gI_0/k}$ the speed of sound. In this regime, the intensity distribution exhibits the well-known
Cherenkov interference pattern, or ``ship-wave'' pattern, characteristic
of a supersonic flow past an obstacle
\cite{Gladush2007, Michel2018}.
For $M<1$, by contrast, the denominator is purely imaginary and the intensity distribution is essentially uniform everywhere, except in the vicinity of the obstacle where the intensity is depleted at the scale of the healing length. This corresponds to the dissipationless superfluid regime.

Remarkably, the spin-orbit coupling introduces a purely imaginary correction to the denominator. Consequently, the real part of the pole condition remains unchanged, and the Landau criterion $M=1$ for superfluidity is therefore unaffected by the gauge field. In the following two subsections, we examine this intensity pattern in detail in the two characteristic regimes, $M>1$ and $M<1$.

\subsection{Chiral Cherenkov radiation}

Let us first focus on the regime $M>1$, where scattering from the obstacle is dissipative. For concreteness, from now on we consider a Gaussian-shaped obstacle of radius $R$, $V(\br_\perp)=V_0\exp(-r_\perp^2/2R^2)$, so that $V(\bq)=2\pi R^2 V_0\exp(-\bq^2R^2/2)$.
\begin{figure}
    \centering
    \includegraphics[width=1.0\linewidth]{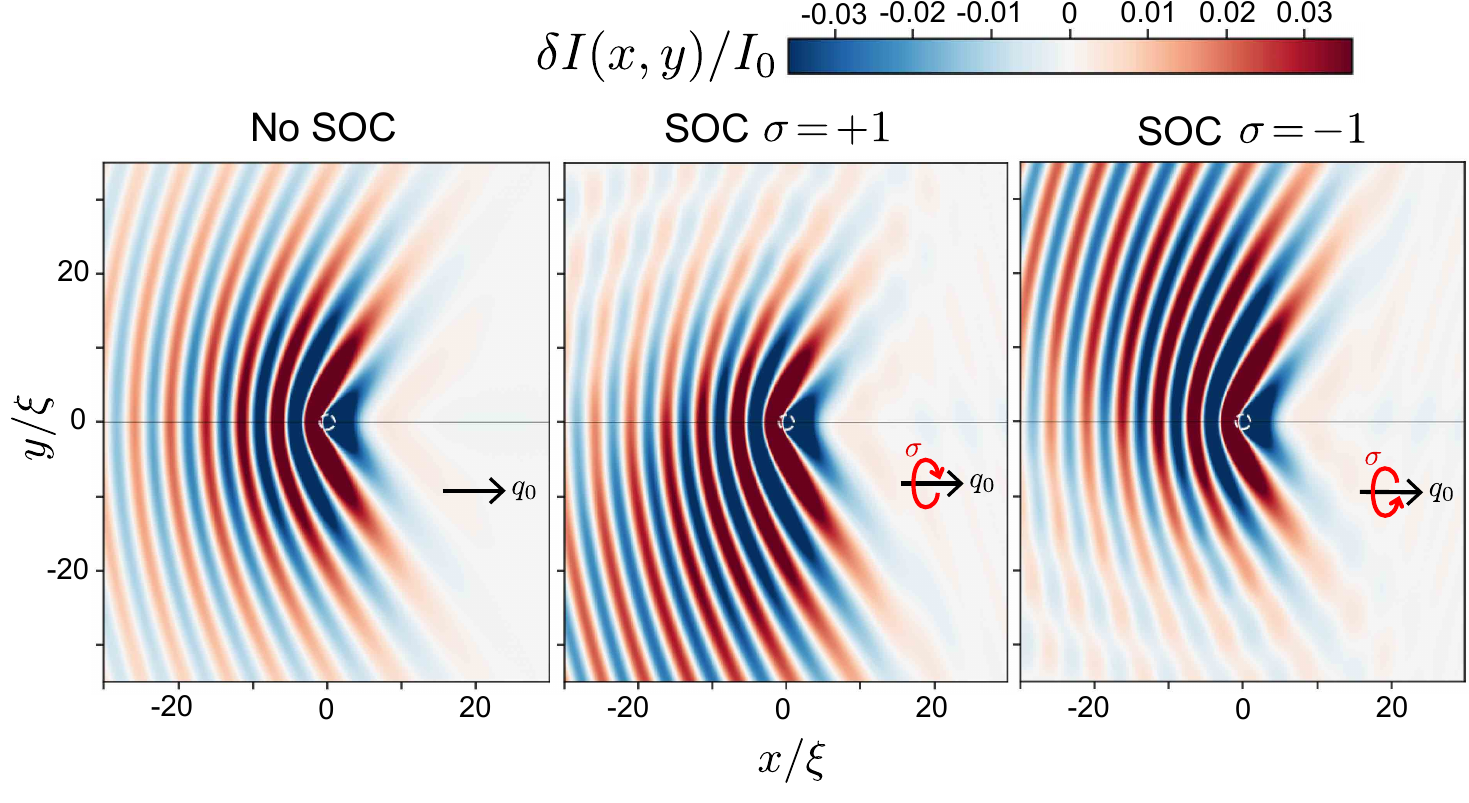}
    \caption{
    Intensity distribution $\delta I(x,y)/I_0$ of a fluid of light flowing past a Gaussian-shaped obstacle at $M=1.36$, above the Landau threshold for superfluidity, in the absence (left panel) and presence of spin-orbit coupling (SOC) (middle panel, $\sigma=1$; right panel, $\sigma=-1$). The intensity distribution is obtained by numerically evaluating Eq.~\eqref{eq:Intensitypattern}. In the absence of the gauge field, it exhibits a characteristic Cherenkov radiation pattern, which becomes chiral in its presence: for $\sigma=1$, the intensity is attenuated for $y>0$ and enhanced for $y<0$, with the opposite behavior for $\sigma=-1$. Parameters: $R/\xi=1$, $q_0\xi=0.96$ ($M=1.36$), $V_0/gI_0=0.2$, $k\xi=2$.
    }
    \label{fig:cherenkov}
\end{figure}
We first show in Fig. \ref{fig:cherenkov} a numerical plot of the intensity pattern \eqref{eq:Intensitypattern} in the regime $M>1$, both in the absence (left panel) and in the presence (middle panel, $\sigma = +1$; right panel, $\sigma=-1$) of the spin-orbit gauge field. Without the spin-orbit term, the pattern reproduces the characteristic Cherenkov waves in the wake of the obstacle. 
In the presence of the (small) spin-orbit term, the Cherenkov radiation becomes \emph{chiral}: the symmetry $\delta I(x,-y)=\delta I(x,y)$ is broken, and, depending on the helicity, one branch of the Mach cone is attenuated while the other is enhanced (with a mirror picture when the  helicity is changed from $+1$ to $-1$). This chirality can be understood from Eq.~\eqref{eq:Intensitypattern}: in the presence of the gauge field, the poles of the integrand acquire a finite imaginary part proportional to the beam helicity $\sigma$ and to $q_y$. 
This helicity-dependent transverse asymmetry is reminiscent of a \emph{spin-Hall-type} response.
We will now show that a similar phenomenon also occurs in the superfluid regime, $M<1$.

\subsection{Chiral response of optical superfluids}

Next, we consider the case $M<1$, for which the two poles of the denominator of Eq.~\eqref{eq:Intensitypattern} are purely imaginary. In this regime, the fluid of light behaves as a superfluid and flows without friction around the obstacle. This manifests itself as an essentially uniform intensity pattern, except in a small region of size $\sim\xi$ around the obstacle, where the intensity is locally depleted.
\begin{figure}
    \includegraphics[width=1.0\linewidth]{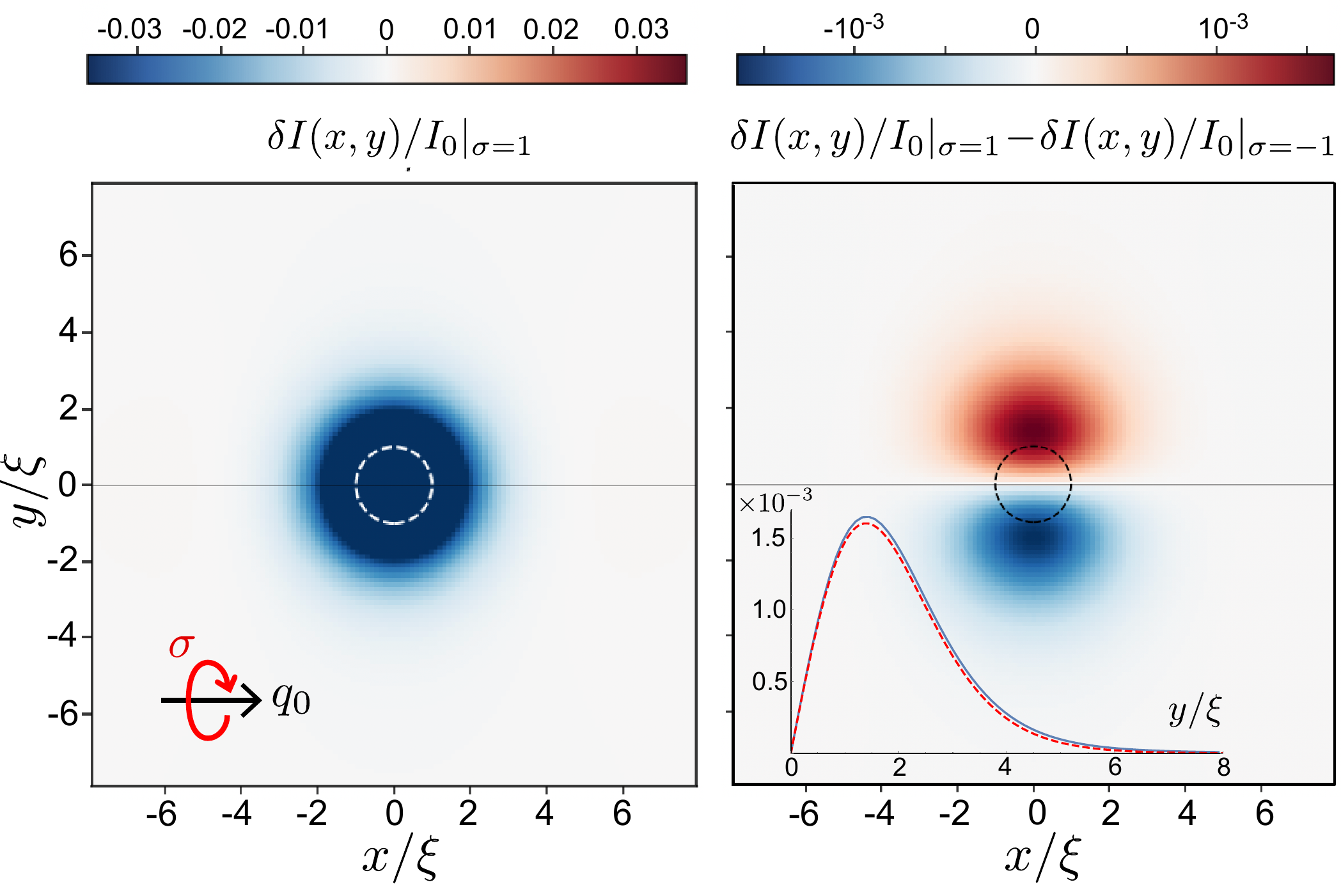}
    \caption{
    Intensity distribution of the optical superflow past a Gaussian-shaped obstacle for $M=0.28$ (the obstacle is indicated as a dashed circle).
    Left: intensity perturbation $\delta I(x,y)/I_0|_{\sigma=1}$ computed numerically from Eq.~\eqref{eq:Intensitypattern}. 
    Although the spin-orbit gauge-field contribution is included, the resulting pattern is dominated by the approximately isotropic depletion around the obstacle.
    Right: difference between the intensity patterns obtained for opposite helicities,
    $\delta I(x,y)/I_0|_{\sigma=1}-\delta I(x,y)/I_0|_{\sigma=-1}$,
    which isolates the contribution that is odd under $\sigma\to-\sigma$ and reveals the dipolar structure associated with the gauge field.
    The inset shows a cut of the intensity difference along the $y>0$ direction ($x=0$): the blue curve corresponds to the exact evaluation of Eq.~\eqref{eq:Intensitypattern}, while the red curve shows the analytical result \eqref{eq:difference} obtained in the regime $M\ll1$.
    Parameters: $R/\xi=1$, $q_0\xi=0.2$ ($M=0.28$), $V_0/gI_0=0.2$, and $k\xi=2$. 
    }
    \label{fig:superflow}
\end{figure}
We show in the left panel of Fig.~\ref{fig:superflow} the intensity distribution computed numerically from Eq.~\eqref{eq:Intensitypattern} for $M\simeq 0.28$. Although this distribution includes the contribution of the spin-orbit coupling, it appears, at first sight, very similar to that expected in the absence of a gauge field \cite{Pavloff2002, Astrakharchik2004, Frisch1992, Pomeau1993, Josserand1999, Pinsker2014, Huynh2024}. However, a clear signature of the gauge field emerges when considering the \emph{difference} between the intensity patterns obtained for opposite helicities, namely
$\delta I(\br_\perp)|_{\sigma=1}-\delta I(\br_\perp)|_{\sigma=-1}$,
shown in the right panel of Fig.~\ref{fig:superflow}. A nonzero residual pattern emerges, taking the form of a dipolar structure: the intensity difference exhibits an enhanced region for $y>0$ and a depleted region for $y<0$, consistent with the breaking of the $y\to-y$ symmetry. Reversing the sign of either $\sigma$ or $q_0$ naturally produces the mirror image of the pattern with respect to the $y=0$ axis. We have verified numerically that this residual dipolar structure persists for any finite value of $M<1$. It is absent only for $M=0$, for which the gauge-field contribution in Eq.~\eqref{eq:Intensitypattern} vanishes. This dipolar structure therefore provides a direct signature of the spin-orbit gauge field in the intensity distribution of an optical superfluid flowing past an obstacle.

To understand the origin of the emerging dipolar structure, it is instructive to evaluate Eq.~\eqref{eq:Intensitypattern} in the small-Mach-number regime, $M\ll1$. Using polar coordinates $q_x=q\cos\theta_q$, $q_y=q\sin\theta_q$, $x=r_\perp\cos\theta$, and $y=r_\perp\sin\theta$, and expanding the denominator of the integrand, we obtain
\begin{align}
\label{eq:expansion}
\frac{ \delta I(\br_\perp)}{I_0}
&\simeq
-\int\! \frac{d^2\bq}{(2\pi)^2}
\frac{4kV(\bq)e^{iq r_\perp\! \cos(\theta_q-\theta)}}{q^2+2/\xi^2}\times
\nonumber \\
&\bigg[
1
+\frac{M^2/\xi^2}{q^2\!+\!2/\xi^2}
+\frac{M^2/\xi^2\cos2\theta_q}{q^2\!+\!2/\xi^2}\\
&+\frac{i\sigma}{(k\xi)^2}\frac{Mq/\xi^2\sin\theta_q}{\sqrt{2}(q^2\!+\!2/\xi^2)}
+\mathcal{O}(M^4)
\bigg]\nonumber.
\end{align}
This expansion naturally organizes the intensity pattern into angular multipole contributions. The first two terms in the brackets yield, after integration over $\theta_q$, the monopole contribution to the intensity pattern, corresponding to the isotropic ``hole'' visible in the left panel of Fig.~\ref{fig:superflow}. The third term yields, after integration over $\theta$, a weak \emph{quadrupolar} contribution $\propto\cos2\theta$ to the intensity. This contribution is barely visible in the left panel of Fig.~\ref{fig:superflow}, but becomes increasingly pronounced at larger $M$. Both the monopole and quadrupole contributions are present in the absence of the gauge field and, at leading order, are unaffected by it.
The most interesting contribution arises from the fourth term in Eq.~\eqref{eq:expansion}. It describes a \emph{dipolar} structure that originates entirely from the gauge field and breaks the $y\leftrightarrow -y$ symmetry. Although this contribution would be difficult to identify directly in practice due to the dominance of the monopole contribution, it can be readily isolated by taking the difference between the intensity patterns obtained for opposite helicities, as done in the right panel of Fig.~\ref{fig:superflow}. With this procedure, the monopole and quadrupole contributions cancel, leaving
\begin{equation}
\label{eq:difference}
\begin{aligned}
&\frac{\delta I(\br_\perp)|_{\sigma=1}
      -\delta I(\br_\perp)|_{\sigma=-1}}{I_0}
\simeq \\
&\quad \quad \frac{2M}{\sqrt{2}(k\xi)^2}\sin\theta
\int_0^\infty \frac{q\,dq}{2\pi}
\frac{4kV(\bq)J_1(qr_\perp)}
{\xi^2(q^2+2/\xi^2)^2}.
\end{aligned}
\end{equation}
This expression directly describes the dipolar structure visible in the right panel of Fig.~\ref{fig:superflow}. In the inset, we show a numerical cut along $x=0$ ($\theta=\pi/2$) of the intensity-difference distribution, together with the analytical prediction given by Eq.~\eqref{eq:difference}. The excellent agreement is consistent with the relatively small value $M\simeq0.28$ used in the numerical evaluation of Eq.~\eqref{eq:Intensitypattern}.

\subsection{Spin Hall force}

The chirality observed in the intensity pattern around the obstacle also manifests in the force exerted by the fluid on the obstacle.
In the absence of spin-orbit coupling, this ``drag'' force is oriented along the $x$ axis, and is nonzero only when $M>1$  \cite{Pavloff2002, Astrakharchik2004}. In the presence of the gauge field, a novel component of this force appears along the $y$ axis, i.e., in the direction transverse to the fluid's velocity. This ``spin Hall'' force can be determined from the general expression  $\textbf{F}=-\int d^2\br_\perp\delta I(\br_\perp)\nabla_\perp V(\br_\perp)$. Inserting Eq.~\eqref{eq:deltaI_interm} into this expression and projecting onto the $y$ axis, we find:
\begin{align}
F_y\equiv \textbf{F}\cdot\textbf{e}_y=i\frac{I_0}{k}
\int \frac{d^2\bq}{(2\pi)^2}
q_y\, q^2\frac{V(\bq)^2}{\Delta(\bq)}.
\end{align}
This relation immediately indicates why a transverse force appears: due to the spin-orbit gauge field, $\Delta(q_x,-q_y)\ne \Delta(q_x,q_y)$, such that the integrand is no longer $q_y$-antisymmetric and the integral over $q_y$ becomes nonzero.
\begin{figure}
    \centering
    \includegraphics[width=0.9\linewidth]{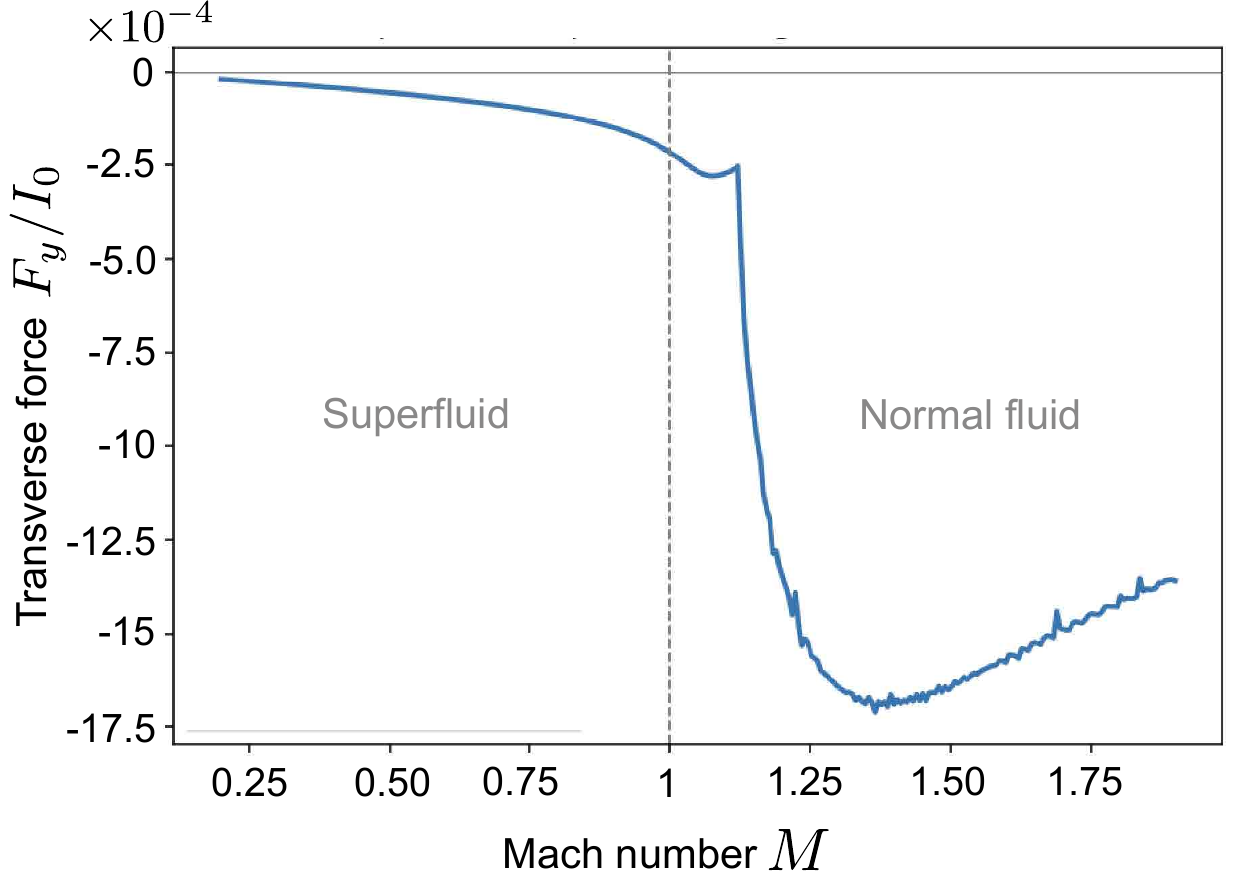}
    \caption{
    Transverse spin Hall force $F_y/I_0$ exerted by the fluid on the obstacle due to the spin-orbit gauge field. The force is small but nonzero in the superfluid regime $M<1$, and becomes much larger in the normal regime $M>1$. The small kink near $M\simeq 1.1$ corresponds to the crossover at which the force changes from reactive to dissipative.
    Parameters: $\sigma=+1$, $R/\xi=1$, $V_0/gI_0=0.05$, and $k\xi=1.5$.
    }
    \label{fig:magnus}
\end{figure}
Figure~\ref{fig:magnus} displays the transverse force $F_y/I_0$ as a function of the Mach number. We observe that the force remains small but nonzero in the superfluid regime and increases abruptly in magnitude near the crossover at $M=1$. We interpret this  behavior as resulting from a change of nature of the force, which turns from purely ``reactive'' in the superfluid regime, i.e. with no radiation emitted to infinity, to mainly ``dissipative'' in the normal-fluid regime.  At large fluid velocities, the obstacle becomes increasingly insensitive to the flow, causing the force to decrease again in magnitude. 

\section{Application 2: spin Hall precession in a focusing fluid}
\label{sec:precession}

In the fluid-flow scenario considered in the previous section, the signatures of the spin-orbit gauge field show up in stationary ($z$-independent) observables. In this section, we discuss a second striking manifestation of the effective gauge field on the fluid of light, in which a non-stationary evolution along $z$ can be exploited to \emph{accumulate} the effect of spin-orbit interactions over the effective propagation time. To this end, rather than considering the stationary regime of a scattering problem as before, we investigate the manifestations of the gauge field in the \emph{oscillatory} dynamics of a \emph{trapped probe}.

\subsection{Weak-probe dynamics in a focusing fluid of light}

To realize this configuration, we now consider a \emph{focusing}  nonlinearity, $g<0$. If the fluid of light is prepared at the medium entrance $z=0$ as a superposition of a strong Gaussian beam and a weak probe, the nonlinear potential $g|\psi|^2\psi$ in Eq.~\eqref{eq:GPEgauge} then essentially acts as an effective harmonic trap for the probe near the center of the background beam. If the probe is initially displaced from the trap center, it will undergo transverse oscillations as a function of $z$, accompanied by a small precessional motion induced by the Lorentz-like force associated with the gauge field. To model this setup and quantitatively test this picture, we seek a solution of  Eq.~\eqref{eq:GPEgauge} in the form
\begin{equation}
\label{eq:linearization}
\psi(\br_\perp,z)=e^{-i\mu z}[\phi_0(\br_\perp)+\delta\phi(\br_\perp,z)].
\end{equation}
In the first term, $\phi_0(\br_\perp)=\sqrt{I_0}\exp(-r_\perp^2/2w_0^2)$ describes the strong Gaussian background beam. 
Note that at leading order, the gauge field term in Eq.~\eqref{eq:GPEgauge} has no effect on this background owing to its radial symmetry: $\boldsymbol{A}\cdot\nabla_\perp\phi_0\propto (\nabla_\perp|\phi_0(\br_\perp)|^2\times \textbf{e}_z)\cdot \nabla_\perp \phi_0(\br_\perp)=0$.
The perturbation $\delta\phi$ describes the probe field, which is assumed to remain weak, $|\delta\phi|\ll|\phi_0|$. In the following we characterize the effect of the effective gauge field on its dynamics.

To realize a Gaussian  background that is indeed stationary 
in a focusing medium, its waist for a given power must be properly chosen such that the kinetic and interaction contributions on the right-hand side of Eq.~\eqref{eq:GPEgauge} compensate each other. Otherwise, the background beam would either diffract or collapse. For a Gaussian beam, this condition can be found from a variational principle \cite{Anderson1979,Zoller1997}, recalled in  Appendix \ref{appendixB}. It leads to the constraint 
$I_0=2/(k|g|w_0^2)$, 
with the chemical potential  given by $\mu=gI_0/4$. We will assume these conditions to be satisfied throughout the remainder of this section. 

To study the dynamics of the probe, we first insert Eq.~\eqref{eq:linearization} into Eq.~\eqref{eq:GPEgauge} and linearize with respect to $\delta\phi$. This yields the Bogoliubov-de-Gennes equation
\begin{align}
\label{eq:BGdGphi0}
i\partial_z\delta\phi=&
\left[
-\frac{(\nabla_\perp-\!ik\boldsymbol{A})^2}{2k}
+\mu+2g|\phi_0(\br_\perp)|^2
\right]\delta\phi\nonumber\\
&+g\phi_0^2(\br_\perp)\delta\phi^*.
\end{align}
Near the beam center, we have
\begin{equation}
   2g|\phi_0(\br_\perp)|^2\simeq
   2gI_0(1-r_\perp^2/w_0^2) =2gI_0+\frac{k\omega_t^2r_\perp^2}{2}
\end{equation}
so that, apart from the anomalous contribution $g\phi_0^2\delta\phi^*$, the evolution of the probe resembles the motion of a particle in a harmonic trap with effective frequency 
\begin{equation}
    \omega_t \equiv\sqrt{\frac{4|g|I_0}{kw_0^2}}=\sqrt{2}|g|I_0,
\end{equation}
where the second equality follows from the stability condition for the background. The vector potential is predominantly determined by the background beam and reads
\begin{equation}
\label{eq:potA}
\boldsymbol{A}=\frac{\sigma g}{2k^2}
(\nabla_\perp |\phi_0(\br_\perp)|^2\times\mathbf{e}_z)=
\frac{\sigma g I_0}{k^2w_0^2}r_\perp \textbf{e}_\theta.
\end{equation}
With this expression for the gauge field, the first line of Eq.~\eqref{eq:BGdGphi0} describes a two-dimensional Schr\"odinger problem analogous to that of a charged particle in a harmonic potential subjected to a uniform magnetic field perpendicular to the plane (the Fock-Darwin problem) \cite{Fock1928, Darwin1931, Baena-Jimenez2026}. 
The problem can be further simplified by noting that the term proportional to $\boldsymbol{A}^2$ in Eq.~\eqref{eq:BGdGphi0} is subleading. We can therefore rewrite  the Bogoliubov-de Gennes equation as
\begin{align}
\begin{aligned}
    \label{eq:nlse_so}
    i\partial_z \delta\phi(\br_\perp,z)&\simeq
    \left(\hat{H}_0-\omega_\text{SO}\hat{L}_z\right)\delta\phi(\br_\perp,z)\\
    &+\hat{\Gamma}_0\,
    \delta\phi^*(\br_\perp,z),
    \end{aligned}
\end{align}
where $\hat{H}_0\equiv -\nabla_\perp^2/2k+k\omega_t^2r_\perp^2/2+\mu+2gI_0$, $\hat\Gamma_0\equiv gI_0+{k\omega_t^2r_\perp^2}/{4}$, $\hat{L}_z\equiv-i\partial_\theta$, 
and we have introduced the characteristic frequency
\begin{equation}
\label{eq:omegasodef}
    \omega_\text{SO}\equiv\frac{\sigma g I_0}{k^2w_0^2}.
\end{equation}
In Eq.~\eqref{eq:nlse_so} the term proportional to $\omega_\text{SO} \hat{L}_z$ generates a rigid rotation of the solution by an angle $\omega_\text{SO} z$ relative to the solution in the absence of gauge field. This is readily seen by introducing the transformation $\delta\phi(\br_\perp,\theta,z)=\varphi(\br_\perp,\Theta,z)$, with $\Theta=\theta+\omega_\text{SO}z$. Substituting this transformation into Eq.~\eqref{eq:nlse_so} eliminates the spin-orbit coupling term, and the  wave function in the rotating frame satisfies
\begin{align}
\begin{aligned}
    \label{eq:BGdG_psi}
    i\partial_z \varphi(\br_\perp,z)=
    \hat{H}_0\varphi(\br_\perp,z)
    +\hat{\Gamma}_0\,
    \varphi^*(\br_\perp,z).
    \end{aligned}
\end{align}
Once this equation has been solved, the probe wave function $\delta\phi$ is obtained by transforming back to the laboratory frame.

In the next subsection, we solve this problem quantitatively for a simple example, illustrating the precession  induced by the spin-orbit gauge field.


\subsection{Spin-orbit-induced precession of a shifted Gaussian probe}

To illustrate the impact of the gauge field on the probe dynamics, we consider a simple example in which a probe beam is prepared at $z=0$ in the form Gaussian wave packet of width $a$ and displaced by a small distance $x_0\ll a$ along the $x$ axis from the center of the trap:
\begin{align}
\begin{aligned}
\label{eq:initialstate}
    \delta\phi(\br_\perp,z=0)&=\sqrt{\delta I_0}
    \exp\Big[-\frac{(x-x_0)^2+y^2}{2a^2}\Big]\\
    &\simeq 
    \sqrt{\delta I_0}\exp\left(-\frac{r_\perp^2}{2a^2}\right)\left(1+\frac{x x_0}{a^2}\right).
\end{aligned}
\end{align}
For simplicity, we choose the width $a=1/\sqrt{k\omega_t}$, which coincides with that of the ground state of the harmonic trap. In the rotating frame, the initial state can then be rewritten as
\begin{align}
\label{eq:psiexpansion}
    \varphi(\br_\perp,0)=\sqrt{\delta I_0\pi} a
    \left[\phi_0(\br_\perp)+\frac{x_0}{\sqrt{2}a}\phi_1(\br_\perp)
    \right],
\end{align}
where $\phi_0(\br_\perp)=\exp(-\br_\perp^2/2a^2)/\sqrt{\pi}a$ and $\phi_1(\br_\perp)=\sqrt{2}r_\perp\cos\Theta\, \phi_0(\br_\perp)/a$  denote, respectively,  the lowest two eigenstates of the two-dimensional harmonic oscillator, with quantum numbers $(n_x,n_y)=(0,0)$ and $(1,0)$.

To describe the subsequent dynamics, we adopt a simple two-mode approximation based on the ansatz
\begin{align}           
\varphi(\br_\perp,z)=
u_0(z)\phi_0(\br_\perp)+u_1(z)\phi_1(\br_\perp),
\end{align}
which is expected to remain valid as long as $x_0/a\ll1$. Substituting this ansatz into the Bogoliubov--de Gennes equation \eqref{eq:BGdG_psi} and projecting onto the two-mode basis, we obtain
\begin{align}
\label{eq:uustart}
i\partial_z
\begin{pmatrix}
u_n\\
u_n^*
\end{pmatrix}=
\begin{pmatrix}
C_n & B_n\\
-B_n&-C_n
\end{pmatrix}
\begin{pmatrix}
u_n\\
u_n^*
\end{pmatrix}
\end{align}
for $n=0,1$, where $C_n=(n+1)\omega_t+\mu+2gI_0$ and $B_n=gI_0+(n+1)\omega_t/4$. Here we have used,  in particular, that $\langle\phi_n|\br_\perp^2|\phi_{n'}\rangle=(n+1)/(k\omega_t) \delta_{nn'}$. The Bogoliubov matrix in Eq. \eqref{eq:uustart} has eigenvalues $\pm\Omega_n$, with $\Omega_n^2=C_n^2-B_n^2$. Using the stability conditions of the background, we find $\Omega_0\simeq 0.375\omega_t$ and $\Omega_1\simeq 0.353\omega_t$ for the Bogoliubov frequencies. The solution for the $u_n$ coefficients then follows as
\begin{align}
\label{eq:solution_u}
    u_n(z)=u_n(0)\left[\cos(\Omega_n z)-i\alpha_n\sin(\Omega_n z)\right]
\end{align}
where $\alpha_n\equiv(C_n+B_n)/\Omega_n$, giving  $\alpha_0\simeq -2.798$ and $\alpha_1\simeq 0.572$. Equations \eqref{eq:psiexpansion} and \eqref{eq:solution_u} constitute the solution of the Bogoliubov problem in the rotating frame. Returning to the laboratory frame, we can finally determine the spatial trajectory of the wave packet in the  harmonic trap from its center-of-mass position, $\br_c(z)\equiv\int d^2\br_\perp\, \br_\perp|\delta\phi|^2/\int d^2\br_\perp|\delta\phi|^2$. We find:
\begin{equation}
\label{eq:trajectory}
\br_c(z)=\tilde{x}_c(z)
\begin{pmatrix}
\cos(\omega_\text{SO}z)\\
-\sin(\omega_\text{SO}z)
\end{pmatrix}_{x,y},
\end{equation}
where 
\begin{align}
   \tilde{x}_c(z)=x_0   \frac{\cos\Omega_0z\cos\Omega_1z+\alpha_0\alpha_1\sin\Omega_0z \sin\Omega_1z}{\cos^2\Omega_0z+\alpha_0^2\sin^2 \Omega_0 z}.
\end{align}
The expression \eqref{eq:trajectory} of the center-of- mass trajectory  deserves a few comments. First, in the absence of the gauge field, we have $\br_c=\tilde x_c \textbf{e}_x$, so that the trajectory remains confined to the $y=0$ plane. This is expected  from the $y\rightarrow -y$ symmetry of the system together with the fact that the initial perturbation is displaced solely along the $x$-direction. The probe field therefore remains even in $y$ throughout the evolution, and the center of mass consequently has no $y$-component. The emergence of a finite $y$ component is thus a direct consequence of  the gauge field, which induces a precession of the trajectory around the $z$ axis.

\begin{figure}
    \centering
    \includegraphics[width=0.8\linewidth]{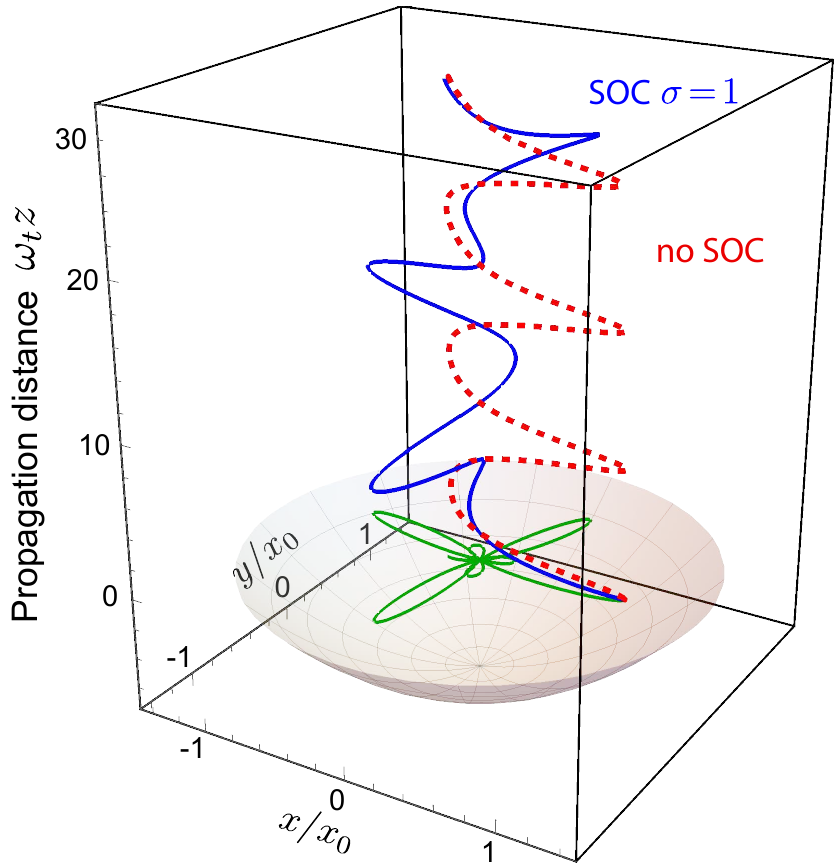}
    \caption{
    Trajectory \eqref{eq:trajectory} of the probe center of mass in the effective trap created by a quasi-harmonic background beam (schematically represented by the light gray shading). The dotted curve shows the trajectory in the absence of the spin-orbit coupling (SOC), while the solid blue curve shows the trajectory when its effect is taken into account. The solid green curve represents the projection of the latter onto the $(x,y)$ plane, highlighting the precession.
    Here we chose $|\omega_\text{SO}/\omega_t|=0.2$ and $\sigma=1$.
    }
    \label{fig:precession}
\end{figure}

The trajectory described by Eq. \eqref{eq:trajectory} is represented in Fig.~\ref{fig:precession} as the effective propagation time $z$ increases. The dashed curve is the oscillatory trajectory of the probe without gauge field, i.e., $\tilde{x}_c(z)$, which stays in the $y=0$ plane. The two Bogoliubov frequencies $\Omega_0$ and $\Omega_1$ are well visible. The solid blue curve shows the trajectory in the presence of gauge field, i.e., including the precession. This precession is better visualized from the projection of the trajectory in the $(x,y)$ plane, which is shown as a solid green curve.
The direction of this precession is determined by the sign of the fluid helicity through Eq.~\eqref{eq:omegasodef}, making it a direct signature of the spin-orbit gauge field. Note that since $\omega_t/\omega_\text{SO}\propto(k\xi)^2\gg1$, the precession is much slower than the harmonic oscillations. Nevertheless, its cumulative effect becomes increasingly visible over longer effective propagation times.\\

\section{Conclusion}

In this work, we have demonstrated that fluids of light deviating from strict paraxiality in Kerr media naturally exhibit an effective gauge field associated with spin-orbit interactions of light. This gauge field is controlled by spatial variations of the nonlinear refractive index and by the helicity of the optical fluid. 
In contrast to previous proposals for synthetic gauge fields for light, which rely on external control such as adiabatic dressing by an auxiliary OAM-carrying pump \cite{Westerberg2016} or dynamic modulation of the refractive index \cite{Fang2012}, the gauge field derived here originates \emph{intrinsically} from the vectorial structure of Maxwell's equations and is self-generated by the fluid, requiring no external intervention or auxiliary field.

Our formalism, encapsulated in Eq.~\eqref{eq:GPEgauge}, provides a framework for exploring the effects of spin-orbit coupling in near-paraxial fluids of light in a broad range of configurations and regimes. In this work, we have illustrated its potential through two examples. First, we have investigated spin-Hall-type shifts and dipolar components in Cherenkov and superfluid patterns produced by a defocusing fluid of light flowing past an obstacle. Second, we have demonstrated Lorentz-like precession of a small probe beam in an effective harmonic trap formed in a focusing fluid.

Our approach, however, is by no means limited to these examples and could be used to investigate a broad range of phenomena, including the interplay between spin-orbit interactions and quantum fluctuations in optical fluids, as well as the role of spin-orbit interactions in the dynamics of optical-fluid mixtures.

\section*{Acknowledgments}

NC is grateful to Quentin Glorieux for relevant comments on the manuscript, and Giovanni Martone for useful discussions. This work was supported by the French National Research Agency under the Grants No. ANR-24-CE30- 6695 FUSIoN and Grant No ANR-23-
PETQ-0001 Dyn1D France 2030.

\appendix

\section{Derivation of leading-order corrections to the paraxial equation}
\label{appendixA}

In this appendix, we provide the detailed steps of the derivation of the beyond-paraxial equation \eqref{Etransversfinal}, starting from Eqs.~\eqref{Etransvers} and \eqref{Ez} for the transverse and longitudinal components of the electric field.

As explained in the main text, the strategy for systematically identifying the relevant nonparaxial terms is to rewrite Eqs.~\eqref{Etransvers} and \eqref{Ez} for the transverse and longitudinal fields in terms of the dimensionless variables introduced in \eqref{eq:dimensionless}. We then perform a controlled expansion of all terms that are not accompanied by either unit prefactors or prefactors proportional to $\eta$. Using Eq. \eqref{eq:dimensionless},
the transverse-field equation \eqref{Etransvers} becomes:
\begin{align}
    i\partial_{\hat z}\Eperphat&=-\nablaperphat^2\Eperphat-\frac{\epsilon^2}{4}\partial^2_{\hat z}\Eperphat{}+|\mathbf{\hat E}_\perp|^2\mathbf{\hat E}_\perp+\eta\mathbf{\hat E}_\perp^2\mathbf{\hat E}^*_\perp\nonumber\\  &+\frac{\epsilon}{2}\nablaperphat\partial_{\hat z}\hat E_z+\nablaperphat(\nablaperphat\cdot\Eperphat)+\frac{i}{\epsilon}\nablaperphat\hat E_z,
    \label{Etransversdimless}
\end{align}
while the dimensionless counterpart of \eqref{Ez} reads
\begin{align}
    &\epsilon^2\nablaperphat^2\hat E_z-i\epsilon\nablaperphat\cdot\Eperphat-\frac{\epsilon^3}{2}\partial_{\hat z}(\nablaperphat\cdot\Eperphat)+\hat E_z\nonumber\\
    &-\epsilon^2|\Eperphat|^2\hat E_z-\epsilon^2\eta\mathbf{\hat E}^2_\perp \hat E_z^*=0.\label{Ezdimless}
\end{align}
In these two expressions, we have already  neglected terms $\propto |E_z|^2$ and $E_z^2$, which are subleading in the $\epsilon\ll1$ expansion. In the next step, we write
\begin{align}
    \hat E_z=i\epsilon\,(\nablaperphat\cdot\Eperphat)+\delta\hat E_z,\qquad \delta\hat E_z=\mathcal{O}(\epsilon^3),
    \label{Ezsplit}
\end{align}
where the leading term $i\epsilon(\nablaperphat\cdot\Eperphat)$ is fixed by the $\mathcal{O}(\epsilon)$ balance in Eq.~\eqref{Ezdimless}, and $\smash{\delta\hat E_z}$ is the correction we still need to determine.  Inserting Eq.~\eqref{Ezsplit} into Eq.~\eqref{Etransversdimless}, we find that the last two terms in Eq.~\eqref{Etransversdimless} simply combine as
\begin{align}   
\label{eq:cancellation}
\nablaperphat(\nablaperphat\cdot\Eperphat)+\frac{i}{\epsilon}\nablaperphat\hat E_z
    =\frac{i}{\epsilon}\nablaperphat\delta\hat E_z,
\end{align}
i.e.\ the $\mathcal{O}(1)$ piece $\nablaperphat(\nablaperphat\cdot\Eperphat)$ cancels {exactly}. With the cancellation \eqref{eq:cancellation}, setting $\epsilon=0$ in Eq.~\eqref{Etransversdimless} immediately yields the paraxial wave equation \eqref{paraxialvectoreq}.

We now determine $\delta\hat E_z$ using the equation for the longitudinal field.
Inserting Eq.~\eqref{Ezsplit} into Eq.~\eqref{Ezdimless} and collecting terms of order $\epsilon^3$ gives
\begin{align}
    \delta\hat E_z
    &=\epsilon^3\left[\frac{1}{2}\partial_{\hat z}(\nablaperphat\cdot\Eperphat)+i|\Eperphat|^2(\nablaperphat\cdot\Eperphat)\right.
    \nonumber\\
    &\left.-i\eta\mathbf{\hat E}^2_\perp (\nablaperphat\cdot\Eperphat^*)-i\nablaperphat^2(\nablaperphat\cdot\Eperphat)\right].
    \label{deltaEz}
\end{align}
Substituting \eqref{deltaEz} together with \eqref{Ezsplit} into \eqref{Etransversdimless}, and keeping terms up to $\mathcal{O}(\epsilon^2)$, we obtain
\begin{align}
 \label{EtransversdimlessEz}
   & i\partial_{\hat z}\Eperphat=-\nablaperphat^2\Eperphat+|\mathbf{\hat E}_\perp|^2\mathbf{\hat E}_\perp+\eta\mathbf{\hat E}_\perp^2\mathbf{\hat E}^*_\perp\nonumber
   -\frac{\epsilon^2}{4}\partial_{\hat z}^2\Eperphat{}\\
    &+i\epsilon^2\partial_{\hat z}\nablaperphat(\nablaperphat\cdot\Eperphat)-\epsilon^2\nablaperphat[|\mathbf{\hat E}_\perp|^2(\nablaperphat\cdot\Eperphat)]\\
    &\!\!\!\!\phantom{\frac{1}{1}}
    +\epsilon^2\eta\nablaperphat[\mathbf{\hat E}_\perp^2(\nablaperphat\cdot\Eperphat^*)]+\epsilon^2\nablaperphat\nablaperphat^2(\nablaperphat\cdot\Eperphat).
    \nonumber
\end{align}

The right-hand side of equation \eqref{EtransversdimlessEz} still contains the $z$-derivative $\partial_{\hat z}(\nablaperphat\cdot\Eperphat)$, which must be re-expressed in terms of $\Eperphat$ alone. 
Taking the divergence of the paraxial equation \eqref{paraxialvectoreq} directly gives, at this order,
\begin{align}
    \partial_{\hat z}(\nablaperphat\cdot\Eperphat)&=    i\nablaperphat^2(\nablaperphat\cdot\Eperphat)-i\,\nablaperphat\cdot\!(|\Eperphat|^2\Eperphat)\nonumber\\
    &-i\eta\nablaperphat\cdot\!(\Eperphat^2\Eperphat^*)+\mathcal{O}(\epsilon^2),
    \label{divzparaxial}
\end{align}
a scalar identity obtained with a single product-rule expansion. Substituting Eq.~\eqref{divzparaxial} into the  first term in the second line of Eq.~ \eqref{EtransversdimlessEz}, and expanding the last two divergences in Eq.~\eqref{divzparaxial},
one finds that the terms $\nablaperphat\nablaperphat^2(\nablaperphat\cdot\Eperphat)$ and $\nablaperphat\bigl[|\Eperphat|^2(\nablaperphat\cdot\Eperphat)\bigr]$ already present in \eqref{EtransversdimlessEz} cancel exactly against matching pieces generated by \eqref{divzparaxial}, while the term $\nablaperphat\bigl[\Eperphat^2(\nablaperphat\cdot\Eperphat^*)\bigr]$ doubles instead of cancelling. What survives is the final, closed nonparaxial wave equation \eqref{Etransversfinal} of the main text.

\section{Stability condition for a stationary Gaussian beam}
\label{appendixB}

In Sec.~\ref{sec:precession}, we consider, in the absence of a probe, a stationary
solution $\psi(\br_\perp,z)=e^{-i\mu z}\phi_0(\br_\perp)$ of Eq.~\eqref{eq:GPEgauge}
for a focusing fluid of light, with $\phi_0(\br_\perp)=\sqrt{I_0}\exp(-r_\perp^2/2w_0^2)$.
As shown in Refs.~\cite{Chiao1964, Anderson1979, Zoller1997}, such a background solution exists
only for specific values of the intensity $I_0$, waist $w_0$, and chemical potential
$\mu$, which we derive in this appendix.

Inserting $\psi(\br_\perp,z)=e^{-i\mu z}\phi_0(\br_\perp)$ into Eq.~\eqref{eq:GPEgauge}
and neglecting the non-paraxial terms yields
\begin{equation}
\label{eqformu}
    \mu \phi_0=-\frac{\nabla_\perp^2\phi_0}{2k}
    +g|\phi_0|^2\phi_0.
\end{equation}
This relation coincides with the Euler-Lagrange equation obtained by extremizing the energy
functional
\begin{equation}
    E[\psi]=\int d^2\br_\perp
    \left(\frac{|\nabla_\perp\psi|^2}{2k}+\frac{g}{2}|\psi|^4\right)
\end{equation}
at fixed power $P=\int d^2\br_\perp|\psi|^2$, with $\mu$ playing the role of the
associated Lagrange multiplier: $\delta(E-\mu P)=0$, or equivalently, $\mu=\delta E/\delta P$. While the Gaussian ansatz is
not an exact solution of Eq.~\eqref{eqformu}, this identity still provides the
correct way to extract $\mu$ once $\phi_0$ is optimized variationally: we first
determine, at fixed $P$, the width $w_0$ that extremizes $E$ within the Gaussian
family, and then evaluate $\mu$ on this optimal family.

We first evaluate the energy for the Gaussian ansatz, as a function of $w_0$ and $P$
(using $I_0=P/\pi w_0^2$):
\begin{align}
\begin{aligned}
\label{eq:energyE}
    E(w_0,P)&=\frac{1}{w_0^2}\left(\frac{P}{2k}+\frac{gP^2}{4\pi}\right).
\end{aligned}
\end{align}
At fixed $P$, extremizing over $w_0$, $\partial E/\partial \omega_0|_P=0$, fixes the power to
the critical value
\begin{equation}
    P=P_c=-\frac{2\pi}{kg},\qquad\Leftrightarrow\qquad I_0=\frac{2}{k|g|w_0^2}.\\
\end{equation}
To obtain $\mu$, we then use the relation $\mu =\delta E/\delta P$, which coincides with $\partial E/\partial P|_{\omega_0}$ evaluated at fixed
$P=P_c$, since $\partial E/\partial {\omega_0}=0$ along
the extremal family. From Eq.~\eqref{eq:energyE},
\begin{equation}
    \left.\frac{\partial E}{\partial P}\right|_{w_0}=\frac{1}{w_0^2}\left(\frac{1}{2k}+\frac{gP}{2\pi}\right),
\end{equation}
which, evaluated at $P=P_c$, gives
\begin{equation}
    \mu=-\frac{1}{2kw_0^2}=\frac{gI_0}{4}.
\end{equation}

\end{document}